\documentclass{svproc}
\usepackage{graphicx}%
\usepackage{multirow}%
\usepackage{amsmath,amssymb,amsfonts}%
\usepackage{mathrsfs}%
\usepackage[title]{appendix}%
\usepackage{xcolor}%
\usepackage{textcomp}%
\usepackage{manyfoot}%
\usepackage{booktabs}%
\usepackage{algorithm}%
\usepackage{algorithmicx}%
\usepackage{algpseudocode}%
\usepackage{listings}%
\usepackage{tabularx}
\usepackage{dirtytalk}
\usepackage{comment}
\usepackage{url}

\begin{document}
\mainmatter              % start of a contribution
\title{Dynamics of toxic behavior in the Covid-19 vaccination debate}
% \title{User's behavior in Online Social Networks in controversial contexts}
%
\titlerunning{Dynamics of toxic behavior} 
% \titlerunning{User's behavior in OSNs}  % abbreviated title (for running head)
\author{Azza Bouleimen \inst{1, 2} \and Nicolò Pagan\inst{2} \and
Stefano Cresci\inst{3} \and Aleksandra Urman\inst{2} \and Silvia Giordano\inst{1}}
\authorrunning{Azza Bouleimen et al.} % abbreviated author list (for running head)

\institute{University of Applied Science and Arts of Southern Switzerland (SUPSI), Switzerland,\\
\email{azza.bouleimen@supsi.ch},\\ 
%WWW home page:
%\texttt{https://www.ifi.uzh.ch/en/scg/people/bouleimen.html}
\and
University of Zürich (UZH), Switzerland,
\and
Institute of Informatics and Telematics, National Research Council (IIT-CNR), Italy
}

\maketitle              % typeset the title of the contribution

\begin{abstract}
In this paper, we study the behavior of users on Online Social Networks in the context of Covid-19 vaccines in Italy. We identify two main polarized communities: Provax and Novax. We find that Novax users are more active, more clustered in the network, and share less reliable information compared to the Provax users. On average, Novax are more toxic than Provax. However, starting from June 2021, the Provax became more toxic than the Novax. We show that the change in trend is explained by the aggregation of some contagion effects and the change in the activity level within communities. In fact, we establish that Provax users who increase their intensity of activity after May 2021 are significantly more toxic than the other users, shifting the toxicity up within the Provax community. Our study suggests that users presenting a spiky activity pattern tend to be more toxic.
% Our study highlights noticeable  behavioral differences between opposing groups in controversial online discussions and suggests that users presenting a spiky activity pattern tend to be more toxic.

% The abstract should summarize the contents of the paper
% using at least 70 and at most 150 words. It will be set in 9-point
% font size and be inset 1.0 cm from the right and left margins.
% There will be two blank lines before and after the Abstract. 
%\dots
% We would like to encourage you to list your keywords within
% the abstract section using the \keywords{...} command.
 %, user behavior
\end{abstract}
\section{Introduction}\label{sec1}
Nowadays, Online Social Networks (OSNs) are a space of broad discussions and exchange of ideas capable of influencing public opinion~\cite{anstead2015social} and actions in several domains~\cite{tufekci2017twitter}. Often, the structure of the network reproduces the partisanship of the users in real life. They tend to aggregate in groups of similar stances on a specific topic~\cite{gaines2009typing}. However, OSNs, by the design of their algorithms, are thought of favoring echo chambers and political polarization ~\cite{van2021social}. %\cite{cinelli2021echo}. 
The latter can present harmful consequences on the online discussion and is susceptible to translating into real-world violence~\cite{gallacher2021online}. To address this noxious phenomenon, designing suitable interventions, both on the platform and on the user level, is of paramount importance~\cite{cresci2022personalized}. Consequently, a thorough understanding of the dynamic of user's behavior is particularly relevant~\cite{tardelli2023temporal} especially in times of crises such as pandemics, wars, and important political moments like national elections~\cite{alyukov2023wartime,budak2019happened}.

% Toxicity is among the harmful behavior that can be identified online. It is defined as "rude, disrespectful, or unreasonable comment that is likely to make someone leave a discussion"~\cite{perspective}. 

Some studies highlight differences in online behavior across users from different political spectra or personality traits~\cite{koiranen2022undercurrents,tadesse2018personality}. For instance, in~\cite{koiranen2022undercurrents}, the authors conducted a survey showing, in the case of Finland, that those farthest from the political center are more likely to leverage provocation for online interactions. Additionally, supporters of left-wing parties favor protective behavior unlike right-wing supporters. 
% While in~\cite{tadesse2018personality}, the authors reveal differences in text writing as well as network position between Facebook users with different Big 5 psychological traits. For instance, they found that users high in openness have a tendency to favor high-frequency function words rather than content words. Conscientious users negatively correlate with the words expressing negative emotions. In contrast, they tend to talk less about unhappy subjects. Extroverts have larger networks, however, where their friends often do not know each other as they belong to different groups of people. \\
Other studies, leveraging digital trace data, infer a set of features from user's social media accounts and build machine learning models to classify them according their ethnicity identification or political affiliation for instance~\cite{singh2016behavioral,pennacchiotti2011machine}. These models achieve high performance but they do not provide insights on the user's features that characterize each class which is highly important when it comes to understanding user's behavior online. 
In~\cite{ertan2023political}, the authors study online user polarization during the 2014 Soma disaster in Turkey, revealing that political polarization can hamper the effectiveness of response operations.

The analysis of the existing literature reveals some studies that focus on specific contexts like accidents or disasters~\cite{ertan2023political}, and others that do not focus on a particular event. Still, none of them convey a complete understanding of the user's behavior online and what shapes it. To shed light on this complex phenomenon, a broader and multi-focused approach is needed, so as to build a complete understanding of the complex human dynamic behavior on OSNs. In this context, our present contribution consists of observing user's behavior, in particular toxicity, during the Covid-19 online vaccination discussion in Italy. Toxicity is among the harmful behaviors that can be identified online and that significantly reduces the quality of the conversation. It is defined as a \say{rude, disrespectful, or unreasonable comment that is likely to make someone leave a discussion}~\cite{perspective}. 
% While this approach surely conveys valuable research contributions, it fails in capturing users' behavior during specific important events, usually around controversial discussions and disputed positions. 

Through this study, we aim to get fresh insights into the possible differences between groups of users and the kind of interventions to design in order to entertain a healthy and safe online discussion. We start from the observations made in ~\cite{bouleimen2023reactions} where the authors identified two polarized communities in the network. They observed the toxicity within these two communities and found that one of the communities is on average more toxic than the other over time. However, at a certain point in time, the trend is inverted. We significantly extend the research of~\cite{bouleimen2023reactions} by highlighting important differences in the behavior and structures of the two main communities (Section~\ref{sec:community}). We then analyze the evolution of the toxicity within the communities across time and provide explanations for the observed change in the trend (Section~\ref{sec:toxicity}). Finally, we summarize the findings and draw insights into differences in users’ behavior in the context of controversial discussions online (Section~\ref{sec:conclusion}).

%
% The Introduction section, of referenced text~\cite{bib1} expands on the background of the work . The introduction should not include subheadings.

% Springer Nature does not impose a strict layout as standard. 

\section{Data Processing}\label{data}
We base our study on the VaccinItaly dataset~\cite{pierri2021vaccinitaly}, a collection of $\sim$12 million tweets relative to the Covid-19 pandemic in Italy. The dataset covers the period from Dec. 20\textsuperscript{th} 2020 to Oct. 22\textsuperscript{nd} 2021. The data collection was based on a set of Italian keywords related to vaccines. The choice of the keywords was made in a way that reflects the discussion of both pro-vaccination and anti-vaccination users~\cite{pierri2021vaccinitaly}. Around half of the tweets are retweets, and the other half is almost equally split between original tweets, replies, and quotes. The dataset involves 551,816 unique users, where 86\% of them have less than 20 tweets in the dataset.

To obtain a set of users that is representative of the discussion, we selected a subset that abides by a set of criteria adapted from~\cite{trujillo2022make}. Specifically, we define \textit{core users} those users that: (\textit{i}) have at least 20 tweets in the dataset, and (\textit{ii}) published at least a tweet per week for a consecutive duration of 3 months. This choice allows us to identify 9,278 core users (1.7\%) who are responsible for nearly half of the tweets in the whole dataset.
To obtain the toxicity scores of the tweets, we used Detoxify~\cite{Detoxify}, a neural network model that achieved top scores in multiple Kaggle competitions and that was profitably used in several studies to compute toxicity scores for social media content~\cite{rossetti2023bots,maleki2021applying}. Detoxify includes a multilingual model for non-English texts. For example, it achieved AUC of 89.18\% for the Italian language~\cite{Detoxify}. The model returns a score ranging from 0 (low toxicity) to 1 (high toxicity). 
In addition to toxicity, we also measured the credibility of the shared links in the dataset by resorting to NewsGuard~\cite{newsguard}. NewsGuard is an organization of trained journalists to track data points on news websites which are consequently used to automatically score the credibility of a website. Scores range from 0 (low credibility) to 100 (high credibility). Thanks to this strategy we obtained credibility scores for 52\% of the links shared by the core users.

\section{Network communities}
\label{sec:community}
\renewcommand{\thefootnote}{\fnsymbol{footnote}}
The purpose of the study being to characterize the dynamic of user's behavior on OSNs, we build the retweet network of users (resolution parameter equal to 0.7). It is a directed weighted network where nodes represent the core users and edges represent retweets. The weights of the edges represent the number of times a retweet happened from one core user to another. The obtained network has 8,975 nodes,\footnote{303 core users exclusively retweeted non-core users and were therefore absent from the final graph.} 643,907 weighted edges, and is built based on 2,214,149 retweets.
\vspace{-10pt}
\subsection{Community detection}
\begin{figure}[htbp]
    \centering
    \includegraphics[width=0.75\textwidth]{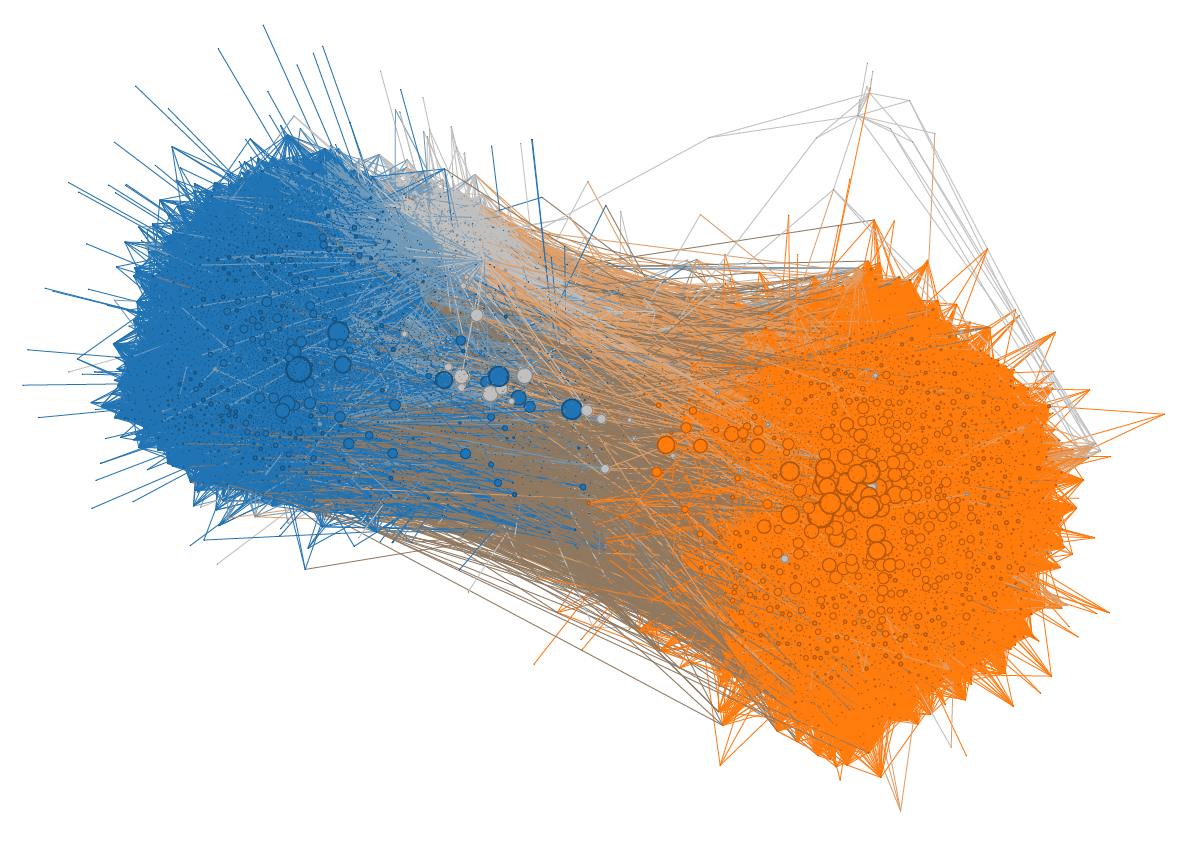}
    \vspace{-15pt}
    \caption{Retweet network of the Covid-19 Italian online debate. The Provax community is blue-colored, the Novax one is orange-colored, and the remaining users are grey-colored.}
    \label{fig:graph}
\end{figure}

We applied the Louvain community detection algorithm~\cite{Blondel_2008} to the retweet network, 
% Initially, we run the algorithm with the standard resolution parameter of 1. We obtained three main communities. However two of these communities were very close in terms of distance, amount of activity, and stance toward the vaccines. We decreased the resolution parameter of the algorithm to 0.7 to favor larger communities. 
obtaining two main communities which gather 87\% of all nodes in the network. The third largest community is constituted by 384 users. The remaining nodes were partitioned into 191 additional communities of much smaller size.
We qualitatively analyzed the tweets of the nodes with the highest authority score in the two main communities~\cite{kleinberg1998authoritative}. This score reflects the tendency of a node to be the source of information in the network. In one community, the nodes with high authority scores tweet content in favor of the vaccines while in the other community, the nodes with high authority scores are against the adoption of vaccines and the government’s measures to contain the spread of the virus. The same observations are found when analyzing the most retweeted tweets in every community or the tweets of the most central users in the two communities. Hence, we deduce that one community is dominated by a \textbf{Pro vaccination} discourse and the other one is dominated by an \textbf{Anti vaccination} discourse. In the following we will refer to these two communities as the \textbf{Provax} community (3,980 nodes) and the \textbf{Novax} community (3,831 nodes). A representation of the retweet network and the communities is shown in Fig.~\ref{fig:graph}. When inspecting the third largest community, and some of the smaller ones, we noticed that they group users favorable for the vaccines or news pages that share reliable information about the vaccines. In the rest of the paper, we will refer to the 192 additional communities as \textbf{Other}.
\vspace{-15pt}
\vspace{-30pt}
% One over the 10 most shared URLs in the Provax community supported the effectiveness of the vaccine. While 4 out of the 10 most shared URLs shared by the Novax community supported their position against the vaccines. 3 of these shared links are likely to be fake news. In fact, we used NewsGuard~\cite{newsguard} to estimate the credibility of the domains of the shared URLs. On average, domains shared within the Provax community have a score of 85 / 100 (0 being not credible at all and 100 very credible) while domains shared within the Novax community have a score of a 48 / 100. The analysis of the hashtags used in both communities gave little indication on the topic of the community.

\subsection{Stability of communities}
In addition to analyzing the network in Fig.~\ref{fig:graph}, obtained by aggregating all data over the whole data collection period, we also built a different network for each month. For each of these networks we then obtained the main communities by means of the Louvain algorithm. Then, for every two consecutive months, we computed the Jaccard similarity between the communities found in the respective networks. The two communities having the highest similarity were matched, assuming that they represent the same community that \say{evolved} from one month to the next one. This dynamic analysis, covering a time period spanning from December 2020 to October 2021, reveals that the majority of the Provax and Novax members remain in their respective community throughout time. The few members that change community, move to other marginal communities. Nevertheless, there is a minority of users that switches between the Provax and Novax communities. Given that the communities are overall stable over time, it is reasonable to study the evolution of the behavior of users over time for the two communities calculated on the retweet network of Fig.~\ref{fig:graph}.
\vspace{-10pt}
\subsection{Differences between Provax and Novax}
\label{sec:stats_communities}
In this section, we discuss some characteristics of the Provax and Novax communities related to their activity, credibility, network structure, and toxicity. In Tab.~\ref{tab:comparison_provax_novax}, we compare most of these characteristics.

\begin{comment}
\begin{table}[htbp]
    \centering
    \makebox[\textwidth]{
        \begin{tabularx}{1.3\textwidth}{@{}>{\centering\arraybackslash}X|>{\centering\arraybackslash}X|>{\centering\arraybackslash}X|>{\centering\arraybackslash}X|>{\centering\arraybackslash}X|>{\centering\arraybackslash}X|>{\centering\arraybackslash}X|>{\centering\arraybackslash}X|>{\centering\arraybackslash}X@{}} \hline
        Community & Number of users & Number of tweets & Number of tweets including URLs & Mean credibility & Number of egdes & Density & Clustering coefficient & Reciprocity \\\hline
        Provax    & 3,980           & 1,684,722        & 256,830                         & 85.86            & 150,152         & 0.009   & 0.14                   & 0.0339      \\\hline
        Novax     & 3,831           & 3,698,329        & 574,807                         & 53.43            & 397,827         & 0.271   & 0.24                   & 0.0665   \\  \hline
        \end{tabularx}
    }
\caption{Observations on the Provax and Novax communities.}
\label{tab:comparison_provax_novax}
\end{table}
\end{comment}

\begin{table}[t]
    \scriptsize
    \centering
    \begin{tabular}{lcccccccccccccccc}
        \toprule
        &&&&&& \textbf{tweets} &&&&&&&& \textbf{clustering} && \\
        && \textbf{users} && \textbf{tweets} && \textbf{w/ URLs} && \textbf{credibility} && \textbf{egdes} && \textbf{density} && \textbf{coefficient} && \textbf{reciprocity} \\
        \midrule
        Provax  && 3,980 && 1,684,722 && 256,830 && 85.86 && 150,152 && 0.009 && 0.14 && 0.0339 \\
        Novax   && 3,831 && 3,698,329 && 574,807 && 53.43 && 397,827 && 0.271 && 0.24 && 0.0665 \\
        \bottomrule
    \end{tabular}
\caption{Observations on the Provax and Novax communities.}
\label{tab:comparison_provax_novax}
\end{table}

% \begin{table}[htbp]
% \centering
% \makebox[\textwidth]{\begin{tabularx}{1.6\textwidth}{@{}>{\centering\arraybackslash}X|>{\centering\arraybackslash}X|>{\centering\arraybackslash}X|>{\centering\arraybackslash}X|>{\centering\arraybackslash}X|>{\centering\arraybackslash}X|>{\centering\arraybackslash}X|>{\centering\arraybackslash}X|>{\centering\arraybackslash}X|>{\centering\arraybackslash}X|>{\centering\arraybackslash}X@{}} 

% \hline
% Community & Number of users & Number of tweets & Number of edges & Average tweets per user & Clustering coefficient & Density & Number of tweets including URLs & Number of tweets including URLs per user & Mean credibility & Reciprocity \\ \hline
% Provax    & 3980      & 1684722    & 150152    & 423.28                  & 0.14                   & 0.009   & 256830                    & 64.53                            & 85.86            & 0.0339      \\ \hline
% Novax     & 3831      & 3698329    & 397827    & 865.37                  & 0.24                   & 0.271   & 574807                    & 150.04                           & 53.43            & 0.0665      \\ \hline
% \end{tabularx}}
% \caption{Observations on the Provax and Novax communities.}
% \label{tab:comparison_provax_novax}
% \end{table}

\vspace{0pt}
In terms of activity, Provax and Novax present different behavior. We notice that, even though there are slightly fewer Novax users, they post twice as much as the Provax users (3,698,329 and 1,684,722 tweets respectively). In fact, the average Novax user posted around 865 tweets while a Provax would post around 423 tweets. When it comes to sharing URLs, both communities have a similar rate of tweets containing URLs which is around 15\%.
%However, Novax users would share on average 2.3 times more tweets containing URLs than a Provax user. 
In addition, the mean credibility of the shared content, as reflected by the NewsGuard scores, is very different between the two communities. Provax have a mean score of 85.86 indicating links from high credibility domains, while Novax have a mean score of 53.43 suggesting content from questionable sources. On a network structure level, the two communities present additional differences. For instance, the Novax community has more than twice the number of edges (retweet ties) of the Provax community. Novax is three times more dense than the Provax one, 1.7 times more clustered, and has two times more reciprocal ties. 

Overall, we conclude that the Novax users are much more active, denser, and more clustered than the Provax users. The users against the adoption of the vaccines are grouped in one main community (Novax) while the ones in favor of them are split into multiple communities with different sizes as seen when analyzing the Other communities in the partition. Moreover, the quality of the information circulating in the Novax community is significantly lower than the quality in Provax one.
%In the following section, we focus on the toxicity level within the two main communities: Provax and Novax.

%\section{Toxicity in communities}
\section{Investigating toxic behaviors}
\label{sec:toxicity}
\begin{figure}[htbp]
    \centering
    \includegraphics[width=\textwidth]{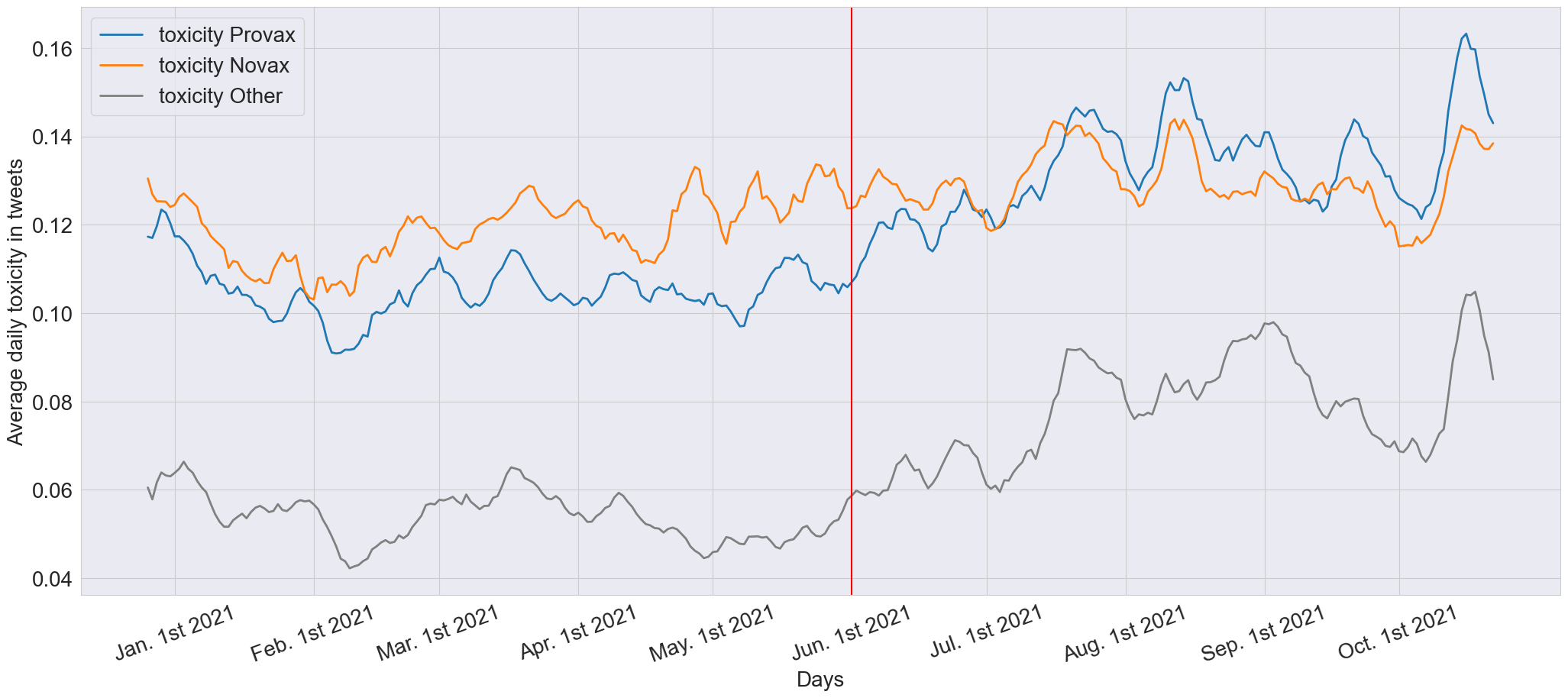}
    \vspace{-20pt}
    \caption{Daily average toxicity in the written text for Provax and Novax communities. A moving average on a 7-day window was applied to the plot.}
    \label{fig:toxicity_communities}
\end{figure}

Next, we study the daily average toxicity of the text written by the users belonging to the Provax and Novax communities. Fig.~\ref{fig:toxicity_communities} shows that, from the beginning of the data collection until June 2021, the toxicity level of the Provax is lower than that of the Novax. Interestingly, this trend is inverted starting from mid-June where the Provax becomes noticeably more toxic than Novax. Meanwhile, the toxicity of Other remains significantly lower than that of both main communities throughout the collection period.
Overall, the toxicity of the Provax, Novax, and Other communities increases across time as shown by the Mann-Kendall test for trends~\cite{kendall1955rank}.
%(Provax: Tau = 0.49, z = 12.85, p $<$ 0.001. Novax: Tau = 0.28, z = 7.45, p $<$ 0.001. Other:  Tau = 0.44, z = 11.55, p $<$ 0.001.)
However, the Provax toxicity rate increases faster than the Novax one. To test the significance of the observed difference we ran a CUSUM test~\cite{CUSUM}, finding that even though the difference is small, it is highly significant (\textit{p}-value $< 0.001$). In summary, Fig.~\ref{fig:toxicity_communities} shows that Provax and Novax have a statistically different behavior, characterized by a different trend with respect to the evolution of toxicity within the two communities. Additionally, Fig.~\ref{fig:toxicity_communities} suggests that the two polarized communities (Provax and Novax) tend to present more extreme behavior than the remaining communities. Finally, the overall increase in toxicity across time in Fig.~\ref{fig:toxicity_communities} might entail a possible increase of the toxicity within the individual users. This hypothesis is rejected following the results from the  Mann-Kendall test for trends. In fact, we found that $\sim$86\% of users in the network do not have a trend, $\sim$10\% of users increase toxicity throughout time, and $\sim$4\% decrease it.

In the remainder of this section, we formulate and test the two following hypotheses to explain the change in the toxicity trends of the Provax and Novax communities observed in Fig.~\ref{fig:toxicity_communities}:
\begin{itemize}
    \item Hypothesis 1 (\textbf{H1}): An increase in the interaction level between Provax and Novax happened around May -- June 2021, which led to a contagion effect.
    \item Hypothesis 2 (\textbf{H2}): A change in activity within the communities happened around May -- June 2021, such that the most active users after that period are more toxic than average.
\end{itemize}
\vspace{-20pt}
\vspace{-15pt}

%\subsection*{Change in toxicity trend}
\subsection{Testing hypothesis H1}
We compute the daily interaction rate between the Novax and Provax communities. Specifically, we consider that user \textit{A} interacted with user \textit{B} if \textit{A} retweeted, quoted, or replied to \textit{B}. 
%Consequently, the daily interaction rate is the number of retweets, quotes, or replies that happened from a user in the Provax (resp. Novax) community to a user in the Novax (resp. Provax) community divided by the total number of tweets (original tweets, retweets, quotes, and replies) posted by the Provax (resp. Novax) users on that day. 
\begin{figure}[htbp]
    \centering
    \includegraphics[width=\textwidth]{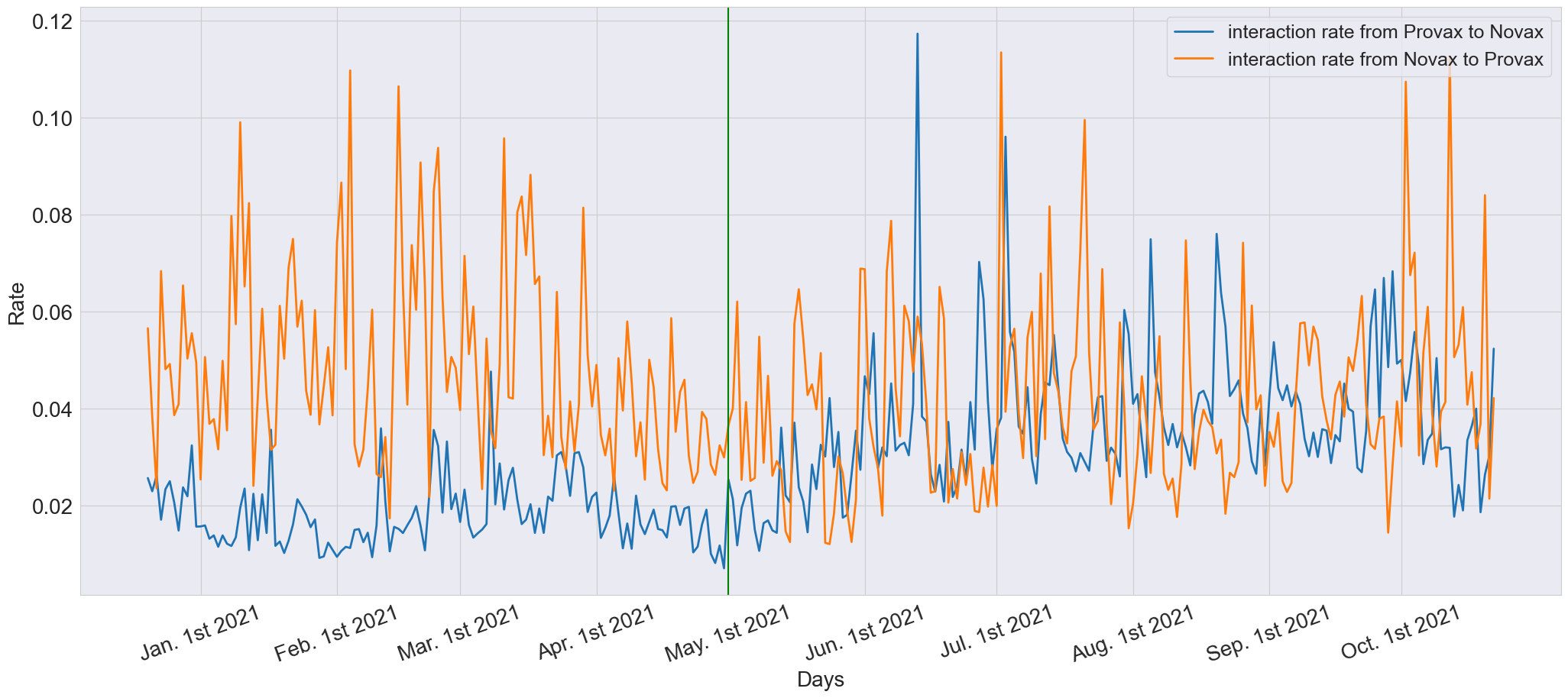}
    \vspace{-20pt}
    \caption{Daily interaction rate from the Provax to the Novax (in blue) and from the Novax to the Provax (in orange).}
    \label{fig:intereactions_provax_novax}
    \setlength{\belowcaptionskip}{0pt}
\end{figure}
% \vspace{-15pt}
Fig.~\ref{fig:intereactions_provax_novax} shows an increase in the rate of interactions from Provax to Novax starting from May 2021, while the interactions from Novax to Provax seem to decrease starting from that period. These observations are confirmed by the Mann-Kendall test of trends (\textit{p}-value $<$ 0.05). In addition, we calculated the same interaction rates for the most toxic users defined as the ones whose toxicity belongs to the last quartile in every community. We found that, overall, the most toxic users in both communities have higher interaction rates with the opposite community. Note that, after May, this rate increases for the most toxic Provax, and decreases for the most toxic Novax. More precisely, \textit{before May 2021}, the most toxic Provax users have on average a 6.5 times higher rate of interactions with Novax than the rest of the Provax. During the same period, the most toxic Novax users have a 2.8 times higher rate of interaction with Provax than the rest of the Novax. Whereas, \textit{after May 2021}, the gap in interaction rates between the most toxic Provax and the rest of the Provax doubles. In fact, the most toxic Provax users become $\sim$13 times more likely to interact with Novax, compared to the rest of the Provax users. This observation is not valid for the Novax users after May 2021: the most toxic Novax users are on average two times more likely to interact with Provax, compared to the rest of the Novax.

Considering the aforementioned observations and the trend in Fig.~\ref{fig:intereactions_provax_novax}, we can conclude that for the Provax, over time, there is an increase in the rate of interaction with the Novax, and that the most toxic Provax users are the ones who are more likely to interact with the Novax. This likelihood increases even more after May, when we start noticing the change in the toxicity trends in Fig.~\ref{fig:toxicity_communities}. Therefore, a possible contagion effect could have occurred from Novax to Provax, which would explain the increase in toxicity observed within the Provax starting from June. However, the amount of Provax-Novax interactions remains limited compared to the overall interactions for every community ($\sim$3\% for the Provax and $\sim$5\% for the Novax). Consequently, in spite of the existence of a contagion effect, it is unlikely that this alone could motivate the change in the toxicity trend shown in Fig.~\ref{fig:toxicity_communities}. We thus conclude that Hypothesis \textbf{H1} cannot be fully retained and other possible explanations should be explored. %This will be the topic of the following paragraph.

\subsection{Testing hypothesis H2}
\begin{figure}[htbp]
    \centering
    \includegraphics[width=\textwidth]{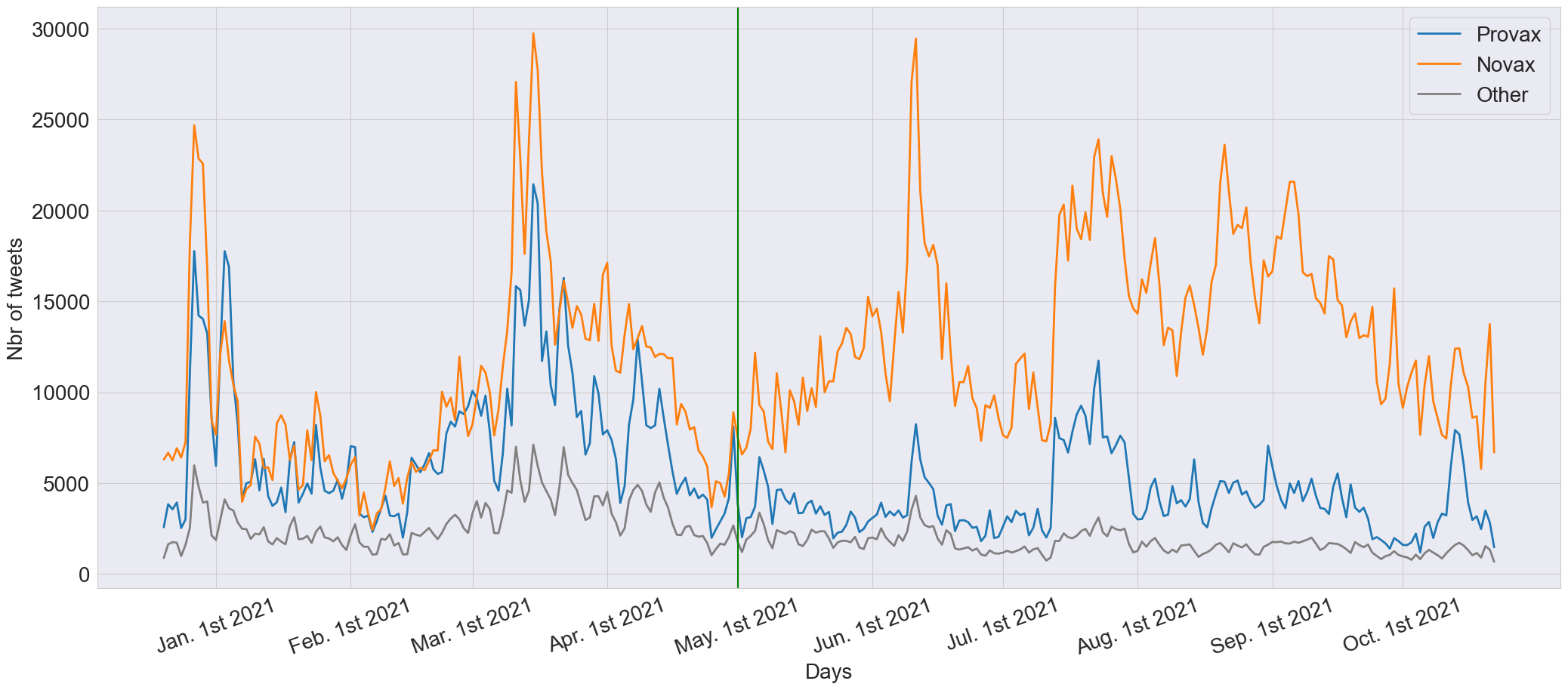}
    
    \vspace{-12pt}
    \caption{Number of tweets posted per day for the Provax, Novax, and Other communities.}
    \label{fig:activity_communities}
\end{figure}

\begin{figure}[htbp]
    \centering
    \includegraphics[width=\textwidth]{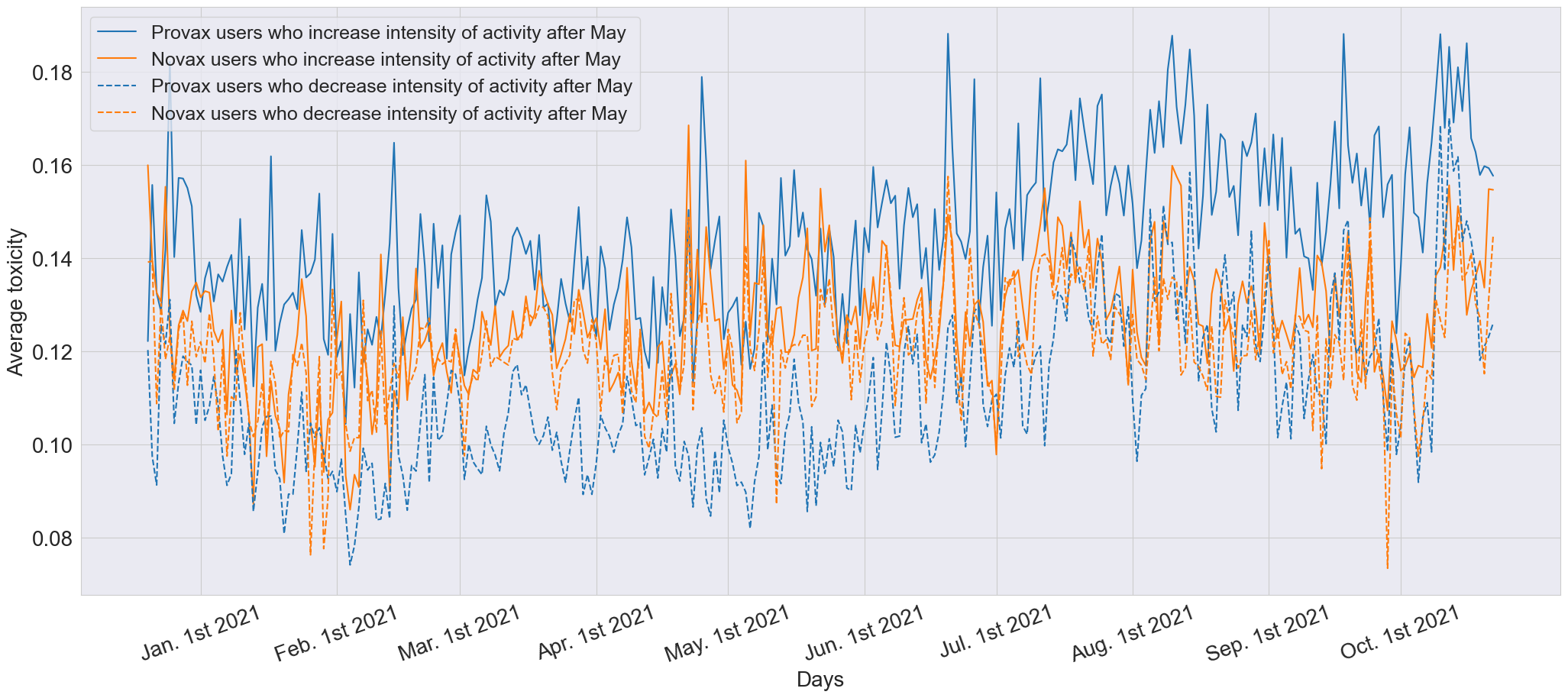}
    \vspace{-20pt}
    \caption{Average of daily toxicity of the Provax and Novax subgroups.}
    \label{fig:toxicity_subgroups}
\end{figure}

To investigate the possible impact of a change in activity on the toxicity of communities, we plot in Fig.~\ref{fig:activity_communities} the number of tweets posted per day for the Provax, Novax, and Other communities. We notice that the activity levels of the Provax and Novax are similar until the end of April when we see an increase in the activity of the Novax and a decrease in the activity of the Provax. This made the difference in tweeting between the two communities important starting in May. Throughout the whole collection period, the activity of the Other is inferior to the ones of the Provax and the Novax and slowly decreases across time.
\vspace{-5pt}
\vspace{-10pt}
Based on these observations, we split the users within the Provax and Novax into two subgroups: users that increased activity (in terms of \textit{number of tweets}) after May 2021 and users that decreased activity after May 2021. We plotted the toxicity evolution for these subgroups for both Provax and Novax communities. There was no statistically significant difference between each pair of subgroups. Then, we decided to measure the activity of users differently. Instead of measuring their activity by counting the number of tweets they posted before and after May 2021, we compute the number of tweets posted by every user divided by the number of days during which that user was active. This measure reflects the activity \textit{intensity} of the user and their tendency to have spikes in their tweeting pattern. With this split, only 18\% of the Provax users increase their activity intensity after May 2021 while 62\% of the Novax do so. The evolution of toxicity of the newly defined subgroups is plotted in Fig.~\ref{fig:toxicity_subgroups}.
% \vspace{-4pt}
% \vspace{-15pt}

In Fig.~\ref{fig:toxicity_subgroups}, we can see that the Provax users who increase the intensity of activity are significantly more toxic than the ones who decrease the intensity. The difference is not that clear between the Novax users who increase and decrease intensity after May 2021. All four trends in the plot are increasing (Mann-Kendall p-value $<$ 0.001). In Tab.~\ref{tab:toxicity_subgroups} we present the mean toxicity scores for the four subgroups. In fact, we can see that on average Provax users that increase intensity after May are more toxic than the other Provax users (0.15 vs 0.11). In addition, their toxicity increase of 0.03 after May, while the Provax users that decrease intensity have their toxicity increase of 0.02. Meanwhile, the Novax subgroups do not have any significant difference between the two corresponding subgroups before and after May 2021.
\vspace{-10pt}
\begin{table}[htbp]
% \makebox[\textwidth]{\begin{tabularx}{1.4\textwidth}
\begin{tabularx}{\textwidth} {@{}c|X|X|X|X@{}}
%{@{}{l>{\centering\arraybackslash}X>{\centering\arraybackslash}X>{\centering\arraybackslash}X@{}}}
\hline
{\color[HTML]{000000} \textbf{Group of user}}  & {\color[HTML]{000000} \textbf{Share of users in community}}               & {\color[HTML]{000000} \textbf{Total mean toxicity}} & {\color[HTML]{000000} \textbf{Mean toxicity before May 2021}} & {\color[HTML]{000000} \textbf{Mean toxicity after May 2021}} \\ \hline
{\color[HTML]{000000} Provax increase intensity*} & 
{\color[HTML]{000000} 0.18} & 
{\color[HTML]{000000} 0.15}                         & {\color[HTML]{000000} 0.14}                              & {\color[HTML]{000000} 0.17}                             \\ \hline
{\color[HTML]{000000} Provax decrease intensity*} & 
{\color[HTML]{000000} 0.82} & 
{\color[HTML]{000000} 0.11}                         & {\color[HTML]{000000} 0.1}                               & {\color[HTML]{000000} 0.12}                             \\ \hline
{\color[HTML]{000000} Novax increase intensity*}  & 
{\color[HTML]{000000} 0.62} & 
{\color[HTML]{000000} 0.13}                         & {\color[HTML]{000000} 0.12}                              & {\color[HTML]{000000} 0.13}                             \\ \hline
{\color[HTML]{000000} Novax decrease intensity*}  & 
{\color[HTML]{000000} 0.38} & 
{\color[HTML]{000000} 0.12}                         & {\color[HTML]{000000} 0.12}                              & {\color[HTML]{000000} 0.12}                             \\ \hline

\end{tabularx}

\caption{Mean toxicity of the Provax and Novax subgroups. * Provax / Novax users who increase / decrease activity intensity after May 2021.}
\label{tab:toxicity_subgroups}
\end{table}
\vspace{-20pt}
Fig.~\ref{fig:toxicity_subgroups} and Tab.~\ref{tab:toxicity_subgroups} support that Provax users who increase intensity after May 2021 are more toxic than the ones who decrease the intensity of their activity. Therefore, we accept Hypothesis \textbf{H2}. This observation is likely to explain the important increase of toxicity within the Provax community that exceeds the one of the Novax in Fig.~\ref{fig:toxicity_communities}. In fact, given the small difference in toxicity between the Novax users who increase and decrease intensity after May 2021, the overall increase of toxicity within the Novax is not as pronounced as the one observed within the Provax community.

Overall, the change in the trend between Provax and Novax communities observed around June is most likely to result from the aggregation of both effects of contagion (\textbf{H1}) and changes in activity within communities (\textbf{H2}). Provax and Novax had different evolutions of activity, characterized by different toxicity levels of the users that increased intensity after May 2021. 
%Consequently, the increase in toxicity with the communities we observe in Fig.~\ref{fig:toxicity_communities} does not imply an increase in the toxicity of the users individually, in opposition to what the plot might suggest. In fact, according to the Mann-Kendall test for trends, we found that $\sim$86\% of users in the network do not have a trend, $\sim$10\% of users increase toxicity throughout time, and $\sim$4\% decrease it.
% \vspace{-30pt}
\section{Discussion and conclusions}
\label{sec:conclusion}
In this work, we studied the behavior of users on social media in the context of controversial topics. From a dataset on the Covid-19 vaccines discussion in Italy, we identified 9,278 core users and built the corresponding retweet network. Leveraging the Louvain community detection algorithm, we identified two main communities: Provax and Novax. The remaining communities in the partition are much smaller but are primarily in favor of the vaccination campaign in Italy. The analysis of the communities revealed that the communities are stable over the whole period covered by the dataset. 

The Novax users are more active and share less reliable information compared to the Provax users. They form groups that are denser and more clustered than the Provax. Moreover, while most of the users against the adoption of the Covid-19 vaccine belong to one main community (Novax), the users in favor of the vaccines are spread across several communities in the network. This suggests that users against the Covid-19 vaccination are more engaged in the discussion,  more clustered together, and have a higher potential for coordination than users in favor of the vaccination. 

Measuring the toxicity within the network, we found that the overall toxicity increases over time. On a community level, Provax and Novax are significantly more toxic than the remaining smaller communities. This suggests, in compliance with other research~\cite{lee2020effects,strandberg2019discussions},  that more polarized communities tend to get more extreme. In addition, we found that, on average, Novax are more toxic than the Provax. However, starting from June 2021, the Provax community become more toxic than the Provax one, suggesting a possible increase in the toxicity of the Provax users. Going more in depth, we rejected this hypothesis as most of the users do not present any trend in the evolution of their toxicity over time. Alternatively, we found that the change in the trend observed is mainly caused by the fact that the overall activity of Provax decreases after May 2021. Yet, the Provax users who increase the intensity of their activity after that date are the ones more toxic on average, driving the community's average toxicity up. This phenomenon is exacerbated by a possible small contagion effect that happens from the Novax to the Provax. In fact, the interaction rate from the Provax to the Novax increases starting from May 2021.

Our work has several implications. First, the differences in the observed behavior between Provax and Novax highlight the complex interplay between users' opinions and their collective behavior. We suggest it is necessary to further explore whether similar observations can be made in the communities divided by other opinion cleavages and, if so, examine what determines the observed differences. Second, even if the toxicity within the communities increases over time, this does not translate into an increase in the toxicity of the individual. The drivers of the change in the toxicity of the user are still unknown so far. However, measuring the activity intensity of the users revealed that an increase in intensity is correlated with a higher toxicity level. Users that present higher spikes in activity patterns are potentially more toxic members of the network. This spiky activity pattern might indicate a behavior that is rather triggered by an external event in particular than the expression of a steady continuous involvement in the online discussion. 

Possible future research directions include studying the reaction of users to specific events and understanding the reasons behind the decrease in activity of Provax after May 2021.

\section*{Acknowledgments}
This work is partially funded by the Swiss National Science Foundation (SNSF) via the SINERGIA project CARISMA (grant CRSII5\_209250), https://carisma-project.org.

\bibliographystyle{spmpsci} % We choose the "plain" reference style
\bibliography{refs} % Entries are in the refs.bib file
\end{document}